\documentclass[12pt]{article}

\usepackage{latexsym}
\usepackage{graphicx}
\usepackage{multicol,multirow}
\usepackage{amsmath,amssymb,amsfonts}
\usepackage{mathrsfs}
\usepackage{amsthm}
\usepackage{rotating}
\usepackage{appendix}
\usepackage{ifpdf}
\usepackage[T1]{fontenc}
\usepackage{times}
\usepackage{newtxmath}
\usepackage{textcomp}%
\usepackage{xcolor}%
\usepackage{hyperref}
\usepackage{lipsum}
\usepackage{subcaption}

\newcommand{\bR}{\mathbf{R}}
\newcommand{\bV}{\mathbf{V}}
\usepackage[margin=2cm, includefoot]{geometry}

\numberwithin{equation}{section}
\numberwithin{thm}{section} % important bit

\newcommand{\bfs}[1]{$\boldsymbol{#1}$}
\newcommand{\ba}{\mathbf{a}}

\usepackage{algorithm}
\usepackage{algpseudocode}

\usepackage{comment}

\usepackage{braket}

\usepackage{tabularx,booktabs}
\newcolumntype{C}[1]{>{\centering\arraybackslash}m{#1}}
\usepackage[labelfont=bf,format=plain,justification=raggedright,singlelinecheck=false]{caption}
\newcommand{\datasetPlot}[2]{
\begin{figure}[h]
		\begin{subfigure}[b]{0.45\textwidth}
			\centering
			\includegraphics[width=0.9\textwidth]{raw_#1.png}
			\caption{Example of experimental images}
		\end{subfigure}
		\hfill
		\begin{subfigure}[b]{0.45\textwidth}
			\centering
			\includegraphics[width=0.9\textwidth]{ca_#1.png}
			\caption{Corresponding enhanced images}
		\end{subfigure}
	
			\begin{subfigure}[b]{0.45\textwidth}
		\centering
		\includegraphics[width=0.9\textwidth]{vol_#1.png}
		\caption{Reconstructed volume }
	\end{subfigure}
	\hfill
	\begin{subfigure}[b]{0.45\textwidth}
		\centering
		\includegraphics[width=0.9\textwidth]{fsc_#1.png}
		\caption{FSC curve of the structure of panel (c)}
	\end{subfigure}
		\captionsetup{justification=centering}
		\caption{#2}
\end{figure} \FloatBarrier}

\usepackage[section]{placeins}

\usepackage{placeins}

\usepackage{multirow}

\newcommand{\nimages}{N}
\newcommand{\nca}{N_c}
\newcommand{\nneighbours}{K}
\newcommand{\ncoeffs}{N_{\text{co}}}
\newcommand{\nangles}{N_{\theta}}

\begin{document}

\title{Signal enhancement for two-dimensional cryo-EM data processing}
\date{}

\author{Guy Sharon, Yoel Shkolnisky, and Tamir Bendory}

\maketitle

%\author[1]{Guy Sharon}
%\author[2]{Yoel Shkolnisky}
%\author[1]{Tamir Bendory}

%\address[1]{\orgdiv{School of Electrical Engineering}, \orgname{Tel Aviv University}, \orgaddress{\city{Tel Aviv}, \state{Israel}}}

%\address[2]{\orgdiv{School of Mathematical sciences}, \orgname{Tel Aviv University}, \orgaddress{\city{Tel Aviv}, \state{Israel}}}

%\authormark{Sharon et al.}

%\keywords{cryo-electron microscopy, signal enhancement, 2-D classification}

%===========================================================================
\begin{abstract}
	Different tasks in the computational pipeline of single-particle cryo-electron microscopy (cryo-EM) require enhancing the quality of the highly noisy raw images. 
	To this end, we develop an efficient algorithm for signal enhancement of cryo-EM images.
	The enhanced  images can be  used for a variety of downstream tasks, such as   2-D classification, removing 
	uninformative images,  constructing  {ab initio} models, generating  templates for particle picking,  providing  a quick assessment of the data set, dimensionality reduction,  and symmetry detection.
	The algorithm includes built-in quality measures to assess its performance and alleviate the risk of model bias. 
	We demonstrate the effectiveness of
	the proposed algorithm on several experimental data sets. 
	In particular, we show that the quality of the resulting images is high enough to produce {ab initio} models of $\sim 10$~\AA~resolution. 
	The algorithm is accompanied by a publicly available, documented and easy-to-use code. 
\end{abstract}

%======================Introduction========================================
\section{Introduction}
In the past few years, single-particle cryo-electron
microscopy (cryo-EM) has become the state-of-the-art method for resolving the atomic structure and dynamics of biological molecules~\cite{frank2006three,kuhlbrandt2014resolution,bai2015cryo,lyumkis2019challenges,murata2018cryo}.
A cryo-EM experiment results in a large set of images, each corresponding to
a noisy tomographic projection of the  molecule of interest, taken from an unknown viewing direction.
In addition, the electron
doses transmitted by the microscope 
must be kept low to prevent damage to the radiation-sensitive
biological molecules, inducing signal-to-noise ratio (SNR) levels that might be as low as -20dB (i.e., the power
of the noise is 100 times greater than the signal)~\cite{survey}.
The low SNR level is one of the main challenges in processing cryo-EM data sets. 
In particular,  different tasks in the computational pipeline of cryo-EM require enhancing the quality of the highly noisy raw images. Specifically, the high-quality enhanced images can be used as 2-D class averages,  
to remove 
uninformative images (e.g., pure noise images, contamination), to construct  {ab initio} models based on the common-lines property~\cite{singer2011three,greenberg2017common}, 
 as  templates for particle picking, to provide a quick assessment of the particles, for dimensionality reduction, and for symmetry detection~\cite{survey,singer2020computational}.

In this paper, we propose a new signal enhancement  algorithm that quickly produces multiple enhanced images that represent different viewing directions of the molecule of interest.
The algorithm begins by performing steerable principal component analysis (sPCA) that reduces  the dimensionality of the data and allows  rotating (steering) the images easily~\cite{pca}.
Next, we randomly choose a subset of images 
and find their nearest neighbors based on (approximately) rotationally invariant operations. Hereafter, we refer to each image and its neighbors as  a class.
Then,  we apply two stages for refining the classes. We first remove   low-quality classes, and then also remove individual images which are inconsistent with their classes.  These stages are based on inspecting the spectra of designed matrices, called synchronization matrices.
The spectra of these matrices (that is, the distribution of their eigenvalues) provide a built-in quality measure to  assess the consistency of each class. This is essential to mitigate the risk in downstream tasks, such as ab initio modeling.
Finally, we run an expectation-maximization (EM) algorithm for each class independently. This step aligns and averages the remaining images in each class, producing high SNR, enhanced images.
The different steps of the algorithm are elaborated in Section~\ref{sec:methods}.

Our algorithm is inspired by and builds on the general scheme of~\cite{zhao2014rotationally}. 
In particular, the authors of~\cite{zhao2014rotationally} suggest finding the rotationally-invariant nearest neighbors of each image  based on the bispectrum: a third-order rotationally-invariant feature~\cite{bendory2017bispectrum}. Then, a high-quality image is produced by aligning all neighbors and averaging. 
While this method works quite well in many cases, the bispectrum inflates the dimensionality of the problem,  boosts the noise level (which is already high in typical data sets) and
does not offer a systematic way to  assess  the performance of the algorithm. 
Section~\ref{sec:methods} elaborates on the differences between ~\cite{zhao2014rotationally} and our proposed algorithm.

Our work also shares similarities with 2-D classification algorithms, which cluster the particle images  and average them to produce high SNR images,  dubbed class averages.
2-D classification is a standard routine  in all contemporary cryo-EM computational pipelines and is mostly used to remove 
uninformative images
that are associated with low-quality class averages and to provide a quick assessment of the particles; our algorithm can be used for those tasks as well. 
A popular solution to the 2-D classification task, implemented in the software RELION, is based on maximizing the posterior distribution of the classes, while marginalizing over the rotations and translations, using an EM algorithm~\cite{relionem}; we describe this methodology in more detail in Section~\ref{sec:EM}. A large number of class averages leads, however, to high computational complexity and to low-quality results because only a few images are assigned to each class. In addition, EM tends to suffer from the ``rich get richer'' phenomenon: most experimental images would correlate well with, and thus be assigned to, the class averages that enjoy higher SNR. As a result, EM tends to output only a few, informative 
classes~\cite{sorzano2010clustering}. 
We circumvent this phenomenon since  we apply the EM to each class separately.

Another related research thread considers denoising at the image or at the micrograph level. 
One example of  the former  is denoising based on  Wiener filtering~\cite{bhamre2016denoising}.
A popular micrograph denoising technique is TOPAZ~\cite{bepler2020topaz}, which is based on deep learning methods; see also~\cite{palovcak2020enhancing,li2022noise}. 
However, these techniques do not harness the similarity between particle images to suppress the noise, and thus their denoising quality is limited. 
For example, we use the enhanced images (the outputs of our algorithm) to directly construct molecular structures at an acceptable resolution. As far as we know, this has not been done using the said methods.

The paper is organized as follows. 
Section~\ref{sec:methods} outlines our method. 
In Section~\ref{sec:results} we present results on  four experimental data sets. 
We attain high-quality images that can be used to construct ab initio models with resolution between $\sim 10$~\AA~to $\sim 20$~\AA. 
Section~\ref{sec:conclusion} concludes the paper and delineates future work to improve the algorithm. 

%====================== Preprocessing ======================
\section{Method} \label{sec:methods}

This section describes the main steps of the proposed algorithm. A documented Python code is available at \url{https://github.com/TamirBendory/CryoEMSignalEnhancement}.

\subsection{Preprocessing}

The algorithm begins with a few standard preprocessing steps: we apply phase-flipping to approximately correct the effect of the CTF, down-sample the images to a size of $89\times 89$ pixels, and whiten the noise in the images. %, and filter the images with a band-pass filter. 
Then, we further reduce the dimension of the images using sPCA, which learns a steerable, data-driven basis for the data set~\cite{pca}. Under this basis, the $i$-th image can be approximated by a finite expansion 
\begin{equation}
    I_i(\xi,\theta) \approx\sum_{k=-k_{max}}^{k_{max}}\sum_{q=1}^{q_k}a^i_{k,q}\psi_c^{k,q}(\xi,\theta), \quad i=1,\ldots,N,
\end{equation}
where $N$ is the number of images in the data set,  $(\xi,\theta)$ are polar coordinate, $\psi_c^{k,q}$ are the sPCA eigenfunctions,  $a^i_{k,q}$ are the corresponding coefficients, $c=1/2$ is the bandlimit of $I$, and $k_{max}$ and $q_k$ are determined as described in~\cite{pca}.
Remarkably, under this representation, an in-plane rotation translates into a phase shift in the expansion coefficients
\begin{equation}
    I_i(\xi,\theta-\alpha) \approx \sum_{k=-k_{max}}^{k_{max}}\sum_{q=1}^{q_k}a^i_{k,q}e^{-\imath k\alpha}\psi_c^{k,q}(\xi,\theta),
\end{equation}
where $\imath=\sqrt{-1}$, 
and, for real-valued images, a reflection translates into conjugation
\begin{equation}
    I_i(\xi,\pi-\theta) \approx  \sum_{k=-k_{max}}^{k_{max}}\sum_{q=1}^{q_k}\overline{a^i_{k,q}}\psi_c^{k,q}(\xi,\theta).
\end{equation}
The sPCA dramatically reduces the dimensionality of the images. We use 500 sPCA coefficients to represent the images.
Henceforth, with a slight abuse of notation, we refer to the vector of  sPCA coefficients of the $i$-th image $\ba_i:=\left\{a^i_{k,q}\right\}_{k,q}$ as the image.

%====================== Nearest Neighbours Search ======================
\subsection{Nearest neighbor search}
Next, we randomly choose $\nca$ images $I_{r_1},\ldots,I_{r_{\nca}}$ from the data set. 
Then, we find the $\nneighbours$ nearest neighbors of each image. The underlying assumption is that the nearest neighbors arise from  similar viewing directions.
We refer to an image and its $K$ neighbors as a \emph{class}.

The nearest neighbor search is based on  a correlation measure which is approximately 
invariant under in-plane rotations and reflection. 
Let $\Theta$  be a predefined set of   $\nangles$  angles; we typically use $\Theta=\frac{i}{36\pi}, i=0,\ldots,71$ so that $\nangles=72.$ 
 We define the approximately invariant correlation, between two images $\ba_i$ and $\ba_j$, by 
\begin{equation} \label{eq:corr}
    \max_{\theta\in\Theta}[\max(\mathrm{corr}(e^{\imath \boldsymbol{k} \theta}\cdot\boldsymbol{a_i},\boldsymbol{a_j}),\mathrm{corr}(e^{\imath \boldsymbol{k}\theta}\cdot\boldsymbol{a_i},\overline{\boldsymbol{a_j}})],
\end{equation}
where 
\begin{equation} 
   \mathrm{corr}(\boldsymbol{u},\boldsymbol{v}) := \frac{(\boldsymbol{u}-\overline{\boldsymbol{u}})^*(\boldsymbol{v}-\overline{\boldsymbol{v}})}{\sigma_{\boldsymbol{u}}\sigma_{\boldsymbol{v}}},
\end{equation}
and where \bfs{k} is a radial frequency vector,  $\cdot$ denotes element-wise product, and $\sigma_{\boldsymbol a}$  is the standard deviation of a vector $\boldsymbol a$.
The nearest neighbors of the $i$-th image are chosen as the $\nneighbours$  images  with 
 the highest  correlation~\eqref{eq:corr}. 

The nearest neighbor search requires computing $N$ correlations for each of the $\nca$  selected images, resulting in a total of $\nca\nimages$ correlations.
For each image, the correlations can be computed  using a couple of matrix multiplications using established linear algebra libraries. 
The computational complexity of this stage is governed by  multiplication of  matrices of size {$\nangles\times\ncoeffs$} and {$\ncoeffs\times\nimages$}.
For the experiments in Section~\ref{sec:results}, this stage took less than a minute.

Two comments are in order. 
First, the approximately invariant correlation can be, in principle, replaced with  invariant polynomials called  the bispectra: analytical rotationally invariant features~\cite{zhao2014rotationally,bendory2017bispectrum}, or approximately rotationally and translationally invariant features~\cite{bendory2021compactification}.
However, the dimension of the bispectrum far exceeds the dimension of the image, and thus we preferred to use the more direct expression of~\eqref{eq:corr}. 
Second, we choose the $\nca$ images at random in order to cover different viewing directions.
In future work, we hope to replace this random strategy with a deterministic technique that finds a set of images having close neighbors while covering all viewing directions. 

We next describe a method to rank and remove low-quality  classes resulting from our random sampling strategy.
This provides a built-in measure of the quality of the classes, and thus of the enhanced images. 

%====================== Grading the Classes ======================

\subsection{Sorting the classes} \label{grading the classes}
Until now, we have randomly chosen a set of images  $I_{i}, i=1,\ldots.\nca$, and found $\nneighbours$ nearest neighbors per class.
However, since the classes were chosen randomly, it is plausible that some of them will be uninformative in the sense that they will not have close neighbors. 
To this end, we aim  to  rank the classes according to their quality.

We define a good class as a class where all of its members were taken from a similar viewing direction, up to an in-plane rotation and, possibly, a reflection.
For each pair of images in the class, we compute the most likely relative in-plane rotation and reflection; this is a by-product of computing the correlations in~\eqref{eq:corr} so no additional computations are required.
We denote the estimated relative rotation angle between  the $i$-th and $j$-th members of the $k$-th class by $\theta_{i,j}^{(k)}$.
If no reflection is involved, the relative rotation can be represented by a  $2\times 2$ rotation matrix
\begin{equation}
    \boldsymbol{R_{i,j}^{(k)}} = \begin{bmatrix}
    \cos{\theta_{i,j}^{(k)}} & -\sin{\theta_{i,j}^{(k)}} \\[6pt]
    \sin{\theta_{i,j}^{(k)}} & \cos{\theta_{i,j}^{(k)}}
    \end{bmatrix}.
\end{equation} 
If the pair of images are also  reflected, then
\begin{equation}
    \boldsymbol{R_{i,j}^{(k)}} = \begin{bmatrix}
    \cos{\theta_{i,j}^{(k)}} & -\sin{\theta_{i,j}^{(k)}} \\[6pt]
    -\sin{\theta_{i,j}^{(k)}} & -\cos{\theta_{i,j}^{(k)}}
    \end{bmatrix}.
\end{equation} 
We then construct a Hermitian block  matrix of size \bfs{R^{(k)}}$\in\mathbb{R}^{2\nneighbours\times 2\nneighbours}$
\begin{equation} \label{eq:synch_mat}
    \boldsymbol{R^{(k)}} = \begin{bmatrix}
    \boldsymbol{R_{1,1}^{(k)}} & \boldsymbol{R_{1,2}^{(k)}} & \cdots & \boldsymbol{R_{1,\nneighbours}^{(k)}} \\[6pt]
    \boldsymbol{R_{2,1}^{(k)}} & \boldsymbol{R_{2,2}^{(k)}} & \cdots & \boldsymbol{R_{2,\nneighbours}^{(k)}} \\[6pt]
    \vdots              & \vdots              & \ddots & \vdots                  \\[6pt]
    \boldsymbol{R_{\nneighbours,1}^{(k)}} & \boldsymbol{R_{\nneighbours,2}^{(k)}} & \cdots & \boldsymbol{R_{\nneighbours,\nneighbours}^{(k)}}
    \end{bmatrix}.
\end{equation} 
The matrix \bfs{R^{(k)}} is  a synchronization matrix over the dihedral group~\cite{bendory2022dihedral}.
If  indeed all $K$ class members are the same image up to an in-plane rotation and, possibly, a reflection (namely, an element of the dihedral group), then  \bfs{R^{(k)}} is of  rank two. 
In other words, only the  first two largest eigenvalues $\lambda_1^{(k)},\lambda_2^{(k)}$ of \bfs{R^{(k)}} are non-zero.
In practice, since the images were not taken precisely from the same viewing direction, and because of the noise, the matrix is not of rank two.
Therefore, as a measure of  the quality of this class, it is only natural to compute how ``close'' is \bfs{R^{(k)}} to a rank two matrix, namely,
\begin{equation}
		G^{(k)} =  \frac{\lambda_1^{(k)}+\lambda_2^{(k)}}{2\nneighbours},
\end{equation}
where we note that 
 $\sum_{i}\lambda_i^{(k)}=\mathrm{Tr}(\boldsymbol{R^{(k)}})=2\nneighbours$.  
We refer to $G^{(k)}\in[0,1]$ as the grade of the $k$-th class. 
We repeat this procedure for each class, and remove the classes with the lowest grades. 
In practice, we found that removing half of the classes  yields good results.
Since we need to extract the two leading eigenvalues of 
$\nca$ synchronization matrices, 
the typical computational complexity of this stage is $O(\nneighbours^2\nca)$, and the worst-case complexity is $O(\nneighbours^3\nca)$~\cite{cheng2005compression}. 

The empirical distributions of the eigenvalues, or the grades $G^{(k)}$, 
provide a measure to assess the performance of the algorithm from the data.
This is important since the output of the signal enhancement algorithm can be used in downstream  procedures, for example, to construct ab initio models, 
Thus, producing deceptive output may  bias the entire computational pipeline with unpredictable consequences~\cite{henderson2013avoiding}. 

%====================== Sorting the Classes ======================
\subsection{Sorting images within classes}
\label{sec:sorting_images}
After removing classes of low quality, we wish to improve each of the remaining classes by removing inconsistent images.  
We follow the same strategy as before, 
and look for, within each class, a subset of images that are consistent with each other, namely, that form an approximately rank-two synchronization matrix~\eqref{eq:synch_mat}. 

Let $\hat\bR^{(k)}=\bV^{(k)}(\bV^{(k)})^*$ be the best rank-two approximation of $\bR^{(k)}$, 
where the columns of $\bV^{(k)}\in\mathbb{R}^{2K\times 2}$ are  the eigenvectors of 
$\bR^{(k)}$ associated with the two leading eigenvalues. Let $\hat\bR^{(k)}[i,j]$  and $\bR^{(k)}[i,j]$ be the $(i,j)$-th entries of 
$\hat\bR^{(k)}$ and $\bR^{(k)}$, respectively.
To determine whether the $i$-th class member is consistent with its other class members, we compute the average distance between $\hat\bR^{(k)}[i,j]$ and $\bR^{(k)}[i,j]$ for all $j=1,\ldots,\nneighbours$. Namely, the grade of each class member is defined by 
\begin{equation}
    g_i^{(k)} = -\frac{1}{\nneighbours}\sum_{j=1}^{\nneighbours}\left\|\hat\bR^{(k)}[i,j]-\bR^{(k)}[i,j]\right\|_2,    
\end{equation}
where $g_i^{(k)}$ is the grade of the $i$-th member of the $k$-th class. 
We found that producing classes with 300 images, and removing 150 images with the lowest score $g_i^{k}$ (per class)  yields good results.

\subsection{Expectation-maximization} \label{sec:EM}
After pruning out low-quality classes, and inconsistent images within each class, we are ready for the last stage of our algorithm: aligning the images within each class and averaging them to produce a high SNR output image.  
To this end, we apply the expectation-maximization (EM) algorithm that aims to maximize the likelihood function of the observed images.
The EM is applied to the raw images of each class separately (before down-sampling, phase-flipping, etc.)

The EM algorithm assumes that all observed images
are rotated, translated, and noisy versions of a single image; this image is  denoted by $X$ and presents the high SNR image we wish to estimate.  
The generative model of the images within a specific class is given by  
\begin{equation}
	I_i = L_{t_i}X + \varepsilon_i, \quad i=1,\ldots, K,
\end{equation}
 where $t_i$ encodes the unknown rotation and translation of the $i$-th image, 
 $L_{t}$ is a linear rotation and translation operator (may also include the CTF), and $\varepsilon_i$ is an i.i.d.\ Gaussian measurement noise with variance $\sigma^2$. 
 Our goal is to maximize the marginalized log likelihood, which is equal, up to a constant, to
 \begin{equation}\label{eq:likelihood}
\log p(I_1,\ldots,I_K;X) = \sum_{i=1}^K \log \sum_{t_\ell\in\mathcal{T}}p(t_\ell) e^{-\frac{1}{2\sigma^2}\|I_i - L_{t_\ell}X\|},
 \end{equation}
  where $\mathcal{T}$ denotes the set of possible rotations and translations. 
 While optimizing~\eqref{eq:likelihood} is a challenging non-convex problem, EM has been proven to be  an effective technique for cryo-EM images~\cite{sigworth1998maximum, survey}. 
 In particular, we used the implementation of  RELION~\cite{relionem}.  
This implementation does not correct for reflections, 
and thus we correct for reflections before running the EM 
based on solving the synchronization problem in~\eqref{eq:synch_mat} using the standard spectral algorithm~\cite{singer2011angular,bendory2022dihedral}.

As mentioned in the introduction, a popular solution to the 2-D classification task 
(which shares similarities with the signal enhancement problem) is to run EM on the observed images, before clustering 
the images. 
 In this case, the generative model reads
\begin{equation} \label{eq:model_many_classes}
	I_i = L_{t_i}X_{m_i} + \varepsilon_i, \quad i=1,\ldots, N,
\end{equation}
where $X_1,\ldots,X_{\nca}$ 
 are the class averages to be estimated. While the implementation of the EM for~\eqref{eq:model_many_classes} follows the same lines as the EM we use, it tends to output only a few informative classes because of the ``rich get richer'' phenomenon~\cite{sorzano2010clustering}. 
We evade this pitfall by first finding the nearest neighbors of the chosen images,  and running EM on each class separately to optimize~\eqref{eq:likelihood}. 

We mention that the EM can be replaced by alternative computational strategies such as stochastic gradient descent or 
 rotationally and translationally aligning the images, and then averaging them. The latter strategy, used by~\cite{zhao2014rotationally}, is much faster than EM, and thus will significantly accelerate the algorithm, at the cost of image quality.

%===========================================================================

\section{Experimental results} \label{sec:results}

In the following experiments, we produced 3000 classes and kept the best $\nca=1500$ classes according to the method explained in Section~\ref{grading the classes}. 
Each class consists of 300 images, from which only the best $K=150$ images were used to estimate the class average, as explained in Section~\ref{sec:sorting_images}. 
We used the EM implementation of RELION~\cite{relionem} with 7 iterations.
Based on the class averages, we reconstructed {ab initio} models   using the common-lines method implemented in the ASPIRE package~\cite{aspire}.
All data sets were processed using an Intel(R) Xeon(R) Gold 6252 CPU @ 2.10GHz containing 24 cores, and a GeForce RTX 2080 Ti GPU.
The run times  of all stages in the process are provided in Table~\ref{table:time}.
The resolution was computed based on  the Fourier Shell Correlation (FSC) criterion with cut-off of 0.5, where the reference volume was downloaded from EMDB~\cite{lawson2016emdatabank}. 

%============================================
%\subsection{EMD-2660} % 10028
\subsection{EMPIAR 10028} % 10028

We begin with a data set of the Plasmodium falciparum 80S ribosome bound to the anti-protozoan drug emetine, available as  the  \texttt{10028} entry in EMPIAR~\cite{iudin2016empiar} (the corresponding entry in EMDB is \texttt{EMD-2660})~\cite{wong2014cryo}. 
This data set contains 105247 images of size $360\times 360$.

Figure~\ref{fig:2660} shows examples of raw data images,  the corresponding class averages, and a 3-D structure reconstructed using the class averages; the resolution of the reconstructed structure is 10.41~\AA.
%\yoelc{the resolution of the reconstructed structure is 10.41~\AA. }{You should say who is the ground truth reference.}
The nearest neighbor stage took 2.5 minutes, and the overall process took around 75 minutes.
%Figure~\ref{fig:snr_vs_k} shows the SNR of the output images as a function of the number of EM input images. 

\datasetPlot{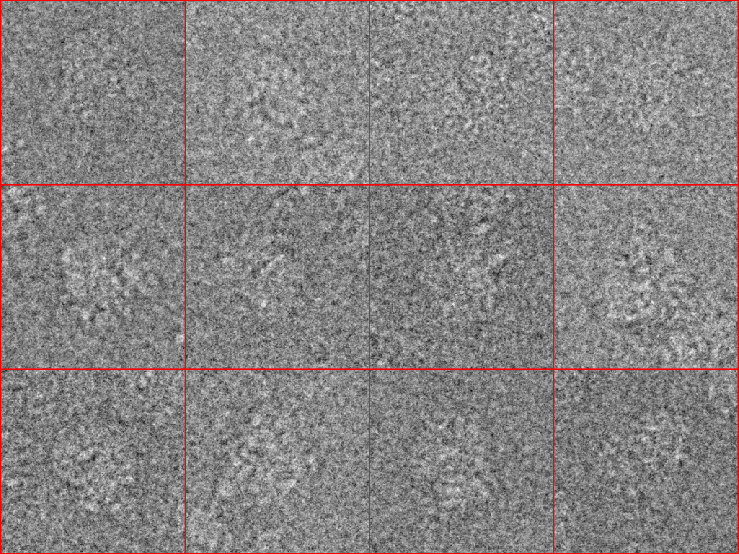}{\texttt{EMPIAR 10028}. The resolution of the reconstructed structure is 10.41\AA\label{fig:2660}}

%\begin{figure}[h]
%	\centering
%	\includegraphics[width=0.4\textwidth]{images/snr_vs_em_num_inputs_2660.png}
%	\caption{\label{fig:snr_vs_k}The average SNR of the enhanced images as a function of the number of input images to the EM for data set \texttt{10028}}	
%\end{figure}

%============================================
%\subsection{EMD-8012} % 10073
\subsection{EMPIAR 10073}

This data set  of the yeast spliceosomal U4/U6.U5 tri-snRNP  is available as the 
\texttt{10073} entry of EMPIAR (the corresponding entry in EMDB is \texttt{EMD-8012})~\cite{nguyen2016cryo}. This data set contains 138840 images of size $380\times 380$. The nearest neighbor search took 2.5 minutes and producing 1500 class averages took roughly 100 minutes.
The results are presented in Figure~\ref{fig:8012}. The resolution of the reconstructed structure is 19.58~\AA.

\datasetPlot{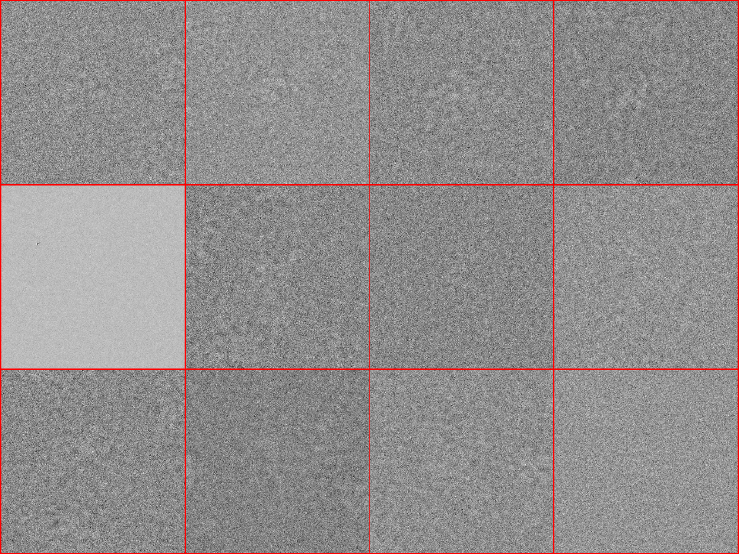} {\texttt{EMPIAR 10073}. The resolution of the reconstructed structure is  19.58\AA\label{fig:8012}}

%============================================
%\subsection{EMD-8511} % 10081
\subsection{EMPIAR 10081} % 10081

This data set of the human HCN1 hyperpolarization-activated cyclic nucleotide-gated ion channel 
is available as the 
\texttt{10081} entry of EMPIAR (the corresponding entry in EMDB is \texttt{EMD-8511})~\cite{lee2017structures}.
 This data set contains 55870 images of size $256\times 256$. 
The nearest neighbor search took less than 3 minutes and producing 1500 high-quality images took roughly~1 hour.
Figure~\ref{fig:8511} shows the results. The resolution of the reconstructed structure is 11.25~\AA.

\datasetPlot{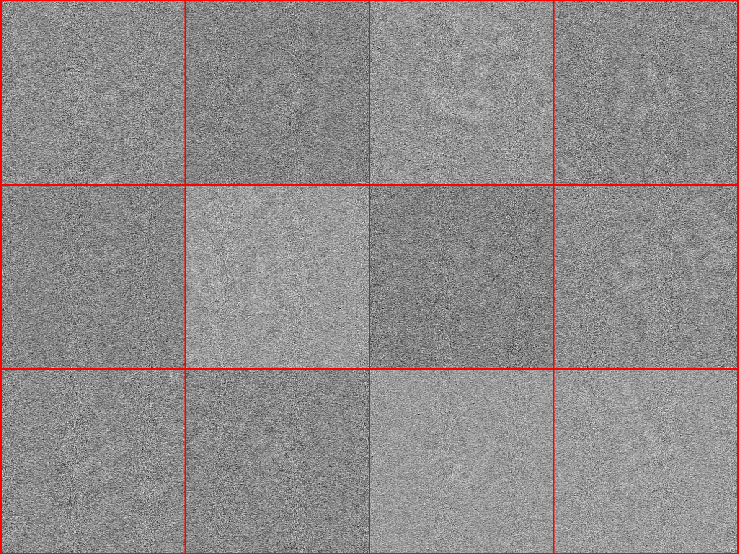}{\texttt{EMPIAR 10081}. The resolution of the reconstructed structure is  11.25\AA\label{fig:8511}}
	
%============================================
%\subsection{EMD-2984} % 10061
\subsection{EMPIAR 10061} % 10061

The next data set represents the structure of beta-galactosidase in complex with a cell-permeant inhibitor,
available as the 
\texttt{10061} entry of EMPIAR (the corresponding entry in EMDB is \texttt{EMD-2984})~\cite{bartesaghi20152}.
This data set contains 41123 images of size $768\times 768$ pixels.
Figure~\ref{fig:2984} shows examples of raw data images, the corresponding enhanced images, and a 3-D structure reconstructed from the enhanced images using the spectral algorithm  implemented in ASPIRE~\cite{rosen2020common}. 
Although the class averages look of good quality, 
the resolution of the reconstructed structure is only 22.63~\AA. The nearest neighbor search took less than 3 minutes and producing 1500 class averages took roughly 2.5 hours.

\datasetPlot{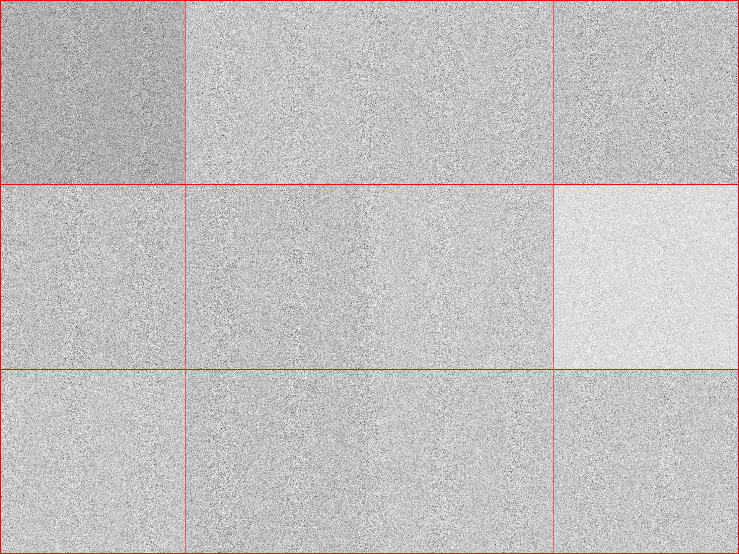}{
	\texttt{EMPIAR 10061}. The resolution of the reconstructed structure is 22.63\AA\label{fig:2984}}

\begin{table}
	\caption{Runtime \label{table:time}} 
\begin{tabular}{ |C{1.5cm}|C{1.5cm}|C{1.9cm}|C{2.3cm}|C{1.1cm}|C{2.2cm}|C{0.8cm}|C{1.0cm}| }
% \hline
% \multicolumn{7}{|c|}{} \\
 \hline
EMPIAR entry & \# images & Dimensions & Preprocessing [sec] & sPCA  [sec] & Nearest neighbor search [sec] & EM [sec] & Total [sec]\\
 \hline
     10028 & 105247 & 360x360 & 1032 & 1036 & 131 & 2299 & 4508 \\ % 10028
     10073 & 138840 & 380x380 & 1735 & 1397 & 155 & 2735 & 6037  \\ % 10073
     10081 & 55870 & 256x256 & 1405 & 698 & 170 & 1149 & 3427  \\ % 10081
     %4905 & 27509 & 250x250 & 203 & 413 & 223 & 2555 & 3399  \\ % 10272
     %6487 & 108544 & 192x192 & 688 & 1095 & 273 & 1694 & 3756  \\ % 10049
     10061 & 41123 & 768x768 & 2405 & 540 & 184 & 5261 & 8413  \\ % 10061
     %7095 & 251851 & 256x256 & 2503 & 2597 & 257 & 2637 & 8013 \\ % 10123"
 \hline
\end{tabular}
\end{table}

%===========================================================================
\section{Discussion} \label{sec:conclusion}

In this paper, we have presented a new algorithm to enhance the quality of  cryo-EM images, which can  be used for various tasks in the computational pipeline of cryo-EM. 
The algorithm is based on~\cite{zhao2014rotationally}, but extends it in several ways, which are crucial to improve its performance and to design built-in quality measures. 
The algorithm is computationally efficient and can be executed on large  experimental data sets. 

For larger data sets, the brute-force nearest neighbor search can be replaced by efficient randomized algorithms~\cite{jones2011randomized} with better asymptotic computational complexity. 
However, for contemporary data sets, the running time of both approaches is comparable. 
Our classification is approximately invariant under in-plane rotations and reflections; see also~\cite{cahill2022group} for a related approach.
While it can be extended to translation invariance by explicitly considering different translations, it will significantly increase the  running time. 
A possible alternative approach would be to  employ polynomials that are approximately invariant under the group of in-plain rotations and translations (namely, the group of rigid motions SE(2))~\cite{bendory2021compactification}.

To improve the quality of the nearest neighbor search, the classes are refined by analyzing the spectrum of synchronization matrices.  
This provides a validation measure that can be computed directly from the data, which is crucial to mitigate the risk of  model bias in downstream tasks, such as ab initio modeling.
We have demonstrated that the enhanced images are of high quality so that they can be used to construct ab initio models.
To cover all viewing angles, we randomly sample the data set. This is clearly not optimal, and we intend to study different deterministic strategies to sample the data. %

%===========================================================================
\section*{Acknowledgment}
T.B. is partially supported by the NSF-BSF award 2019752, the BSF grant no. 2020159, and the ISF grant no. 1924/21. Y.S. was supported by the European Research Council (ERC) under the European Union's Horizon 2020 research and innovation programme (grant
agreement 723991 - CRYOMATH) and by the NIH/NIGMS Award R01GM136780-01.

%===========================================================================
\bibliographystyle{plain}
%\bibliography{refs}

%===========================================================================
%\smallskip
\end{document}